# Ht-Index for Quantifying the Fractal or Scaling Structure of Geographic Features

Bin Jiang[1] and Junjun Yin[2]

[1]Department of Technology and Built Environment, Division of Geomatics
University of Gävle, SE-801 76 Gävle, Sweden
Email: bin.jiang@hig.se

[2]Digital Media Centre, Dublin Institute of Technology, Ireland
Email: yinjunjun@gmail.com

*"Clouds are not spheres, mountains are not cones, coastlines are not circles, and bark is not smooth, nor does lightning travel in a straight line."*
                                                                                            Benoit Mandelbrot

**Abstract**
Although geographic features, such as mountains and coastlines, are fractal, some studies have claimed that the fractal property is not universal. This claim, which is false, is mainly attributed to the strict definition of fractal dimension as a measure or index for characterizing the complexity of fractals. In this paper, we propose an alternative, the ht-index, to quantify the fractal or scaling structure of geographic features. A geographic feature has ht-index *h* if the pattern of far more small things than large ones recurs (*h-1*) times at different scales. The higher the ht-index, the more complex the geographic feature. We conduct three case studies to illustrate how the computed ht-indices capture the complexity of different geographic features. We further discuss how the ht-index is complementary to fractal dimension, and elaborate on a dynamic view behind the ht-index that enables better understanding of geographic forms and processes.

## 1. Introduction
Many geographic features, such as mountains and coastlines, look irregular, wiggly, and rough; they cannot be simply described by their heights and lengths based on Euclidean geometry. For example, the length of a coastline depends on the measuring scale used; the finer the measuring scale, the longer the length, or equivalently, the larger the map scale, the longer the length will be (Richardson 1961) (note the difference between the measuring scale and the map scale; refer to Section 2 for more details). The notion that length is dependent on the map scales is indicated by the straight distribution line in the Richardson plot, in which the x-axis and y-axis represent the logarithms of the measuring scales and lengths respectively (c.f., Figure 1 and the related note there). In theory, the length of a coastline approaches infinity when the measuring scale approaches zero. In practice, a coastline has a finite length because the finest measuring scale is normally larger than zero (e.g., one meter). This phenomenon of undefined lengths of geographic features triggered the development of fractal geometry (Mandelbrot 1967, Mandelbrot 1982), which provided a brilliant new geometry for rough and irregular forms. This is in contrast to Euclidean geometry, which is mainly useful for describing smooth and regular shapes. Fractal geometry has been widely adopted as a very important scientific tool for studying the complexity of nature and society in different disciplines, including economics, physics, biology, and geography. Underlying fractal geometry is the fundamental concept of fractal dimension, which is used to characterize the irregularity or roughness.

The word "fractal" literally means fragmented and irregular, and is a set or pattern for which fractal dimension $D$ has a fractional value that exceeds the topological dimension (Mandelbrot 1982). According to this definition, fractal lines have $D$ greater than 1, and fractal surfaces have $D$ greater than 2. Despite being well-adopted and well-received in the literature, fractal dimension remains one of the most abstract concepts. To making it less abstract, fractal dimension is commonly considered to be the degree of space-filling. A curve with $D$ very close to 1.0 (such as 1.1) behaves much like an ordinary one-dimensional line, but a curve with $D$ very close to 2.0 (such as 1.9) has a very convoluted shape, much like a two-dimensional surface. Unfortunately, this space-filling perspective often creates the incorrect impression that fractal dimension is another measure for density. Essentially, density is a Euclidean concept, while fractal dimension is fractal-based.

Fractal patterns are widespread in nature and society - they appear in geographic features, as well as in our daily lives. Drop an empty beer bottle on cement, and it is likely to break into many pieces. The broken pieces are fractal, for the fragmented pieces have irregular shapes. More importantly, there are far more small pieces than large ones. This notion of far more small things than large ones is also true for geographic features. There are far more small cities than large ones (Zipf 1949), far more short streets than long ones (Carvalho and Penn 2004, Jiang 2009), far more less-connected streets than well-connected ones (Jiang 2007; Jiang 2009), far more low buildings than high ones (Batty et al. 2008), and far more small city blocks than large ones (Lämmer et al. 2006, Jiang and Liu 2012). This list can be extended to include mountains, rivers, lakes, parks, and forests. Geographic features share the same properties of the broken bottle pieces, so they are fractal in essence.

However, some studies in the geographic literature claim that the fractal property is not universal (e.g., Mark and Aronson 1984, Buttenfield 1989, Lam and Quattrochi 1992). This is false! This false claim is mainly due to the strict definition of fractal dimension (c.f., Section 2.2 for more details). We therefore propose an alternative, the ht-index, to quantify the fractal or scaling structure of geographic features. The intention of this study is not to prove that geographic features are fractal, which is obvious, but to capture an aspect not captured by fractal dimension. We will illustrate that the ht-index is complementary to fractal dimension in quantifying the complexity of fractals, or that of geographic features in particular.

The remainder of this paper is structured as follows. Section 2 reviews fractal dimension and further elaborates on its limitations to better motivate the concept of the ht-index. Section 3 defines the ht-index as one plus the recurring times of far more small things than large ones, and illustrates its computation using a workable example of the Koch curve (c.f., Figure 1 and the related note there). Section 4 discusses three case studies involving different geographic features at both the country and city levels to further demonstrate the usefulness and advantages of the ht-index. Section 5 further discusses how the ht-index complements fractal dimension and its implications for better understanding geographic forms and processes. Finally Section 6 draws a conclusion, and points to future work.

**2. Fractal dimension and its limitations**
This section reviews fractal dimension and the closely related concepts of scale and scaling using the Koch curve for illustration. We point out that the concept of scale has a rather different connotation in fractals than it does in geography. We elaborate on the limitations of fractal dimension in order to develop the ht-index.

**2.1 Fractal dimension, scale and scaling**
Fractal dimension, or fractional dimension as initially named by Mandelbrot (1967), is closely related to the notion of self-similarity, i.e., parts look like the whole at multiple scales. Figure 1 and the associated notes illustrate this notion using the famous Koch curve. In this case, self-similarity is defined in a strict sense, because any part is exactly self-similar to the whole curve. Reflected in the Richardson plot, all points are exactly on the distribution line (Figure 1). This rigid definition was relaxed to include self-similarity in a statistical sense (Mandelbrot 1967). Unlike strict self-similarity, statistical self-similarity implies that all points in the Richardson plot are around (rather than on) the distribution line. Fractal dimension is defined as a ratio of the change in detail to the change in scale (Mandelbrot 1982). This is not a simple ratio, but the ratio of logarithms, such as *D = log(N)/log(r)*, where *r* is the measuring scale (or simply change in scale) and *N* is the number of the scale needed to cover the whole fractal pattern or set (or equivalently change in detail). The slope of the distribution line in the Richardson plot (logarithm) is equal to the fractal dimension *D*. The reason not to use the simple ratio is due to the fact that the change in scale r and the change in detail *N* are disproportional. This disproportion implies that decreasing scale *r* by a small amount will dramatically increase the detail *N*, i.e., far more small things than large ones (Figure 1). Taking the Koch curve for example, the scale is decreased by 1/3 every iteration from 1/3, to 1/9 and 1/27 (which is called *scaling factor* or *similarity ratio*), the detail increases four times from 4 to 16 and 64, with D = log(4)/log(3) = 1.26. This indicates what we mean by "far more small things than large ones."

The key concept of fractals is the scale. More specifically, a fractal involves many different scales, ranging from the smallest to the largest. These scales form a scaling hierarchy. In theory, the scale range from the smallest to the largest is infinite for strict fractals such as the Koch curve. In other words, the line segments can be infinitesimally split, so the Koch curve has an infinite length. In practice, many fractals, such as a fern leaf (c.f., Figure 2 for examples), have a finite scale range. For the Koch curve of iteration 3 shown in Figure 1, the smallest line segment is 1/27 given that the curve is generated from a line of one unit. The concepts of scale, scale range, and scaling hierarchy can be seen from the example of Koch curve (Figure 1). The Koch curve of iteration 3 has three scales, including 1/27, 1/9, and 1/3 with respect to 64, 16, and 4 segments, forming a striking scaling hierarchy. Note the scale range is from 1/27 to 1/3.

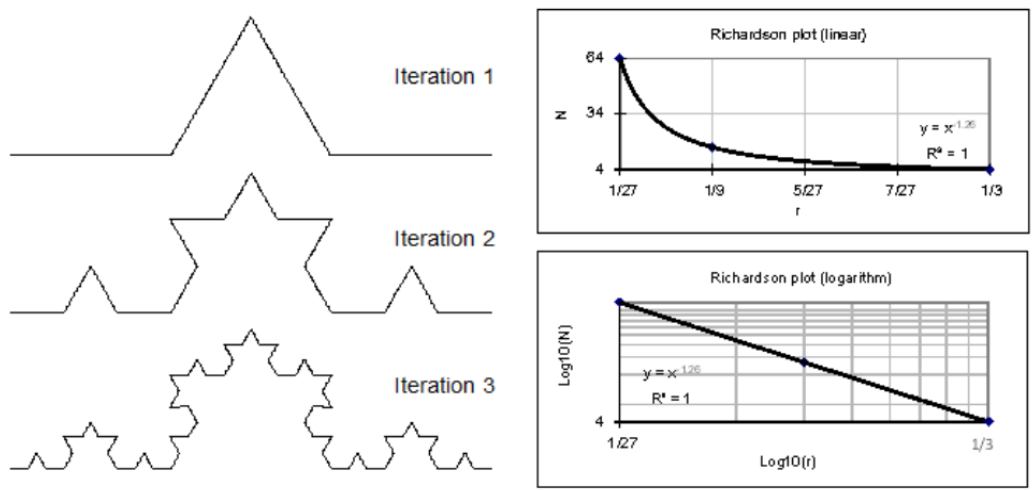

Figure 1: The Koch curves generated from a base line of one unit, and the Richardson plots at both linear and logarithm scales for the Koch curve of iteration 3

(Note 1: Begin with a base line of one unit. Now split the line into thirds and replace the middle third by two sides of an equilateral triangle, leaving you with a broken line of four segments. Now split each of the four segments using this same procedure. Keep splitting each segment and replacing the middle thirds with the two sides of the equilateral triangle repeatedly. In theory, the splitting process continues toward infinity. The result is a Koch curve, invented by the Swedish mathematician Helge von Koch (1870–1924), and belonging to one of many pathological curves, now called fractal curves.

Note 2: The Richardson plot (linear or logarithm) indicates that there are far more short segments than long ones, i.e., 64 short segments and only 20 long ones (c.f., Section 3 for details on how to differentiate short and long segments by the arithmetic mean). Importantly all the segments meet a power law relationship, $y = x^{-1.26}$ , where x is the measuring scale, while y is the length of Koch curve, with R square equal 1. That all three points are exactly on the distribution line in the Richardson plot reflects the strict self-similarity. The slope of the distribution line in the logarithm version of the Richardson plot is in effect the fractal dimension.)

The term *scale* has rather different connotations in fractals than in geography. In fractals, as well as in other sciences such as biology and physics (e.g., Schmidt-Nielsen 1984, Bonner 2006, Bak 1996), the concept of scale is closely related to scaling hierarchy. It is subtly different from the same concept used in geography literature. Scale in geography has three different meanings: cartographic, analysis, and phenomenon (Montello 2001), but none of these three captures the true meaning of scale in fractals. Many geographers tend to take the concept of scale in fractals for granted without noticing the subtle difference. The concept of scale is better seen in the city-size distribution captured by Zipf's law: there are far more small cities than large ones (Zipf 1949). Herein, scale refers to the city sizes in particular, and different scales of cities form a scaling hierarchy.

**2.2 Limitations of fractal dimension**
The definition of fractal dimension requires that changes in scale *r* and detail *N* must meet a power law relationship. Unfortunately, this is too strict for many geographic features. What if the relationship between *r* and *N* is not a power law, but a similar function, such as lognormal or exponential? In this case, the geographic features would be excluded from being fractal or having the associated scaling characteristics. With real-world data, it is sometimes difficult to choose among the three similar functions (power, lognormal and exponential, collectively called heavy-tailed distributions) as the best fit, for they all appear right skewed. Conventionally, a least-squares method was used to find the best-fit function in the Richardson plot. It has recently been found less reliable; instead, a maximum-likelihood method has been suggested (Newman 2005, Clauset et al. 2009). What has been claimed to be a power law under a least-squares method actually may be lognormal, exponential, or a power law with an exponential cutoff. This difficulty in detecting a power law raises a serious issue about the definition of fractal dimension. This is the major reason that some previous studies claimed that the fractal was not universal for geographic features.

This is also the major reason behind relaxing the definition of scaling to include lognormal and exponential for characterizing geographic features (Jiang 2012). It may sound controversial since scaling refers to a power-law relationship in the physics literature. However, the relaxed definition of scaling offers a better way of understanding geographic forms and processes. Geographic features can be less fractal if they have not yet been fully developed. In this case, the relationship between the changes in scale *r* and details *N* might best fit a lognormal or exponential function rather than a power function. A fern leaf is fractal, but it is less so at early stages (Figure 2). Just like the real fern leaf, many phenomena in nature and society are continuously evolving and developing. A power law applies to the stage where phenomena have been fully developed. Before that, phenomena tend to be less like a power law, and more like a lognormal or exponential function.

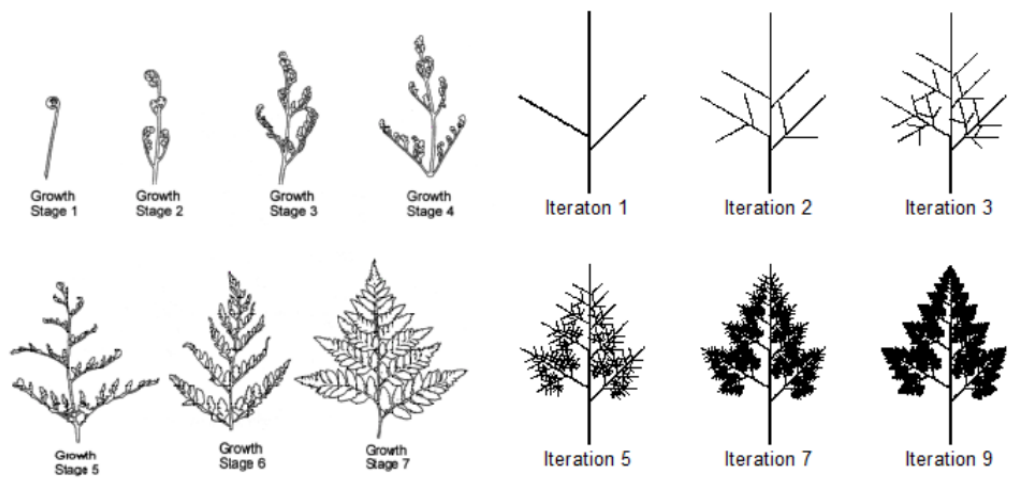

Figure 2: A dynamic view of fractals showing seven growth stages of a real fern leaf (left, Strandberg et al. 1997) and six iterations of a simulated fern leaf (right)

The best instrument to see fractals is the human eye, which can check whether there are far more small things than large ones. Mandelbrot (1982, cited in Taylor 2006) regretted imposing such a strict definition of fractal dimension for seeing fractals: *"For me, the most important instrument…is the eye. It sees similarities before a formula has been created to identify them."* While relying on human eyes, both perspective and scope matter in seeing fractal or scaling patterns of geographic features. For example, street segments or junctions are unlikely to show scaling, but entire streets will show scaling: there are far more less-connected streets than well-connected ones (Jiang et al. 2008). A local view of street networks with a limited number of streets may prevent us from seeing the scaling or fractals.

The fractal dimension is an index for measuring the complexity of fractals, but it does not differentiate the complexity of individual stages or iterations. Figure 2 illustrates a dynamic view of fractals from both the growth and iteration perspectives. The structure of iteration 9 looks far more complex, or fractal, than that of iteration 1. This is because iteration 9 involves more scales than iteration 1, and iteration 1 lacks details, so it can be simply described by Euclidean geometry. From the perspective of filling space, iteration 9 also has far more space filled than that of iteration 1. The same can be said on the left side of Figure 2; stage 7 is far more complex than stage 1. This observation is the same for the Koch curves at different iterations. In this regard, fractal dimension only captures forms, or more precisely fractal forms, not the growth or iteration processes. Fractal dimension does not differentiate structures between more and few scales (e.g., among different iteration stages). This is exactly what the ht-index aims to capture.

**3. Ht-index and nested rank-size plots**
We developed the ht-index to capture how many scales, ranging from the smallest to the largest, form a scaling hierarchy. The ht-index is defined as one plus the recurring times of far more small things than large ones. In other words, a geographic feature has an ht-index of $h$ if the pattern of far more small things than large ones recurs ($h-1$) times at different scales. The higher the ht-index, the more complex the geographic feature. For the sake of simplicity, we will illustrate the concept of the ht-index using the Koch curve of iteration 3 (Figure 1), and indicate how the index can be derived from nested rank-size plots.

The Koch curve has three different scales 1/27, 1/9, and 1/3, so the ht-index is 3. This is fairly straightforward, as we have already known from the above discussion that the three scales ranging from 1/27 to 1/3 form the scaling hierarchy. For illustration purposes, we will assume that we do not know the ht-index, but know the three different sizes of segments: 1/27, 1/9, and 1/3, and their corresponding numbers 64, 16 and 4 (total of 84). Clearly there are far more small segments than large ones. We utilize the arithmetic mean to differentiate between small and large segments. Taking the 84 segments as a whole, the average size (or the first mean) is calculated as follows:

$$m_1 = \frac{4 \times \frac{1}{3} + 16 \times \frac{1}{9} + 64 \times \frac{1}{27}}{84} = 0.07$$

The first mean of 0.07 splits the 84 segments into two unbalanced parts: 20 above the average (ca. 24 percent, a minority), and 64 below the average (ca. 76 percent, a majority). Taking the 20 segments above the first mean, the second mean is computed as follows:

$$m_2 = \frac{4 \times \frac{1}{3} + 16 \times \frac{1}{9}}{20} = 0.16$$

The second mean of 0.16 further splits all the 20 segments into two unbalanced parts: four above the second mean (ca. 20 percent, a minority), and 16 below the mean (ca. 80 percent, a majority). For the remaining four segments in the last head, it makes little sense to further split them, because this would violate the notion of far more small things than large ones.

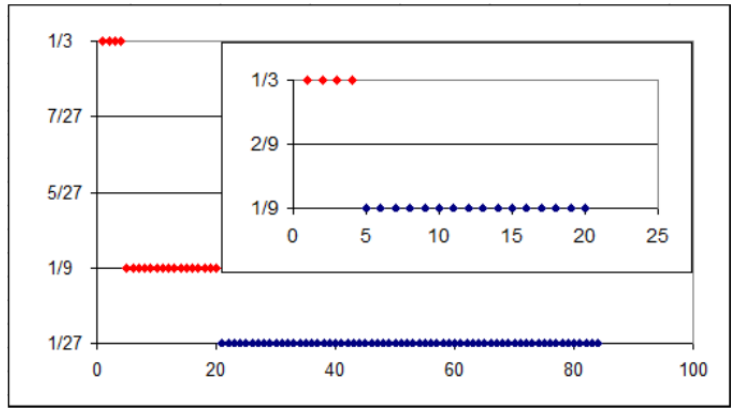

Figure 3: Nested rank-size plots showing a recurring pattern of far more small segments (indicated by blue points) than large segments (indicated by red points)
(Note: The heavy-tailed distribution is plotted iteratively, or in a nested style, for the head part, with the x-axis and y-axis respectively representing the rank and size of the segments from largest to smallest.)

Through the above simple computation, we can see that the pattern of far more small things than large ones occurred two times with the Koch curve. Basically, we ranked all the segments from the largest (with size 1/3) to the smallest (with size 1/27), demonstrating a heavy-tailed distribution with the head and tail respectively representing large and small segments. The head can again be plotted as a heavy-tailed distribution. This is shown in the nested rank-size plots (Figure 3). Based on the ranking and splitting processes illustrated above, or so called head/tail breaks (Jiang 2012), we derived the three hierarchical levels for the Koch curve of iteration 3 (Figure 4).

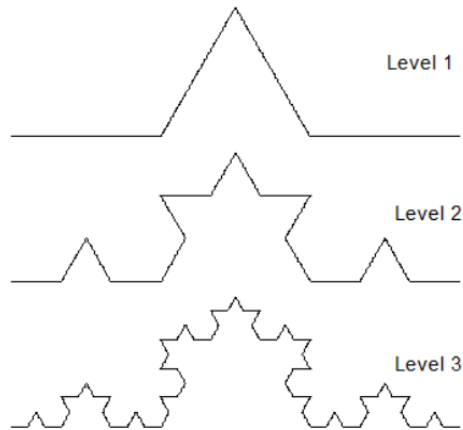

Figure 4: Three scales (ht-index = 3) of the Koch curve of iteration 3 automatically derived by the head/tail breaks
(Note: The three scales or levels look the same as the three iterations in Figure 1, but they are automatically derived from the Koch curve of iteration 3. That is the reason that we label them as levels rather than iterations.)

The nested rank-size plots provide a simple and intuitive approach for examining the recurring pattern of far more small things than large ones. The ht-index itself is relatively straightforward to derive. Two points should be noted with the Koch curve example. First, this example is just for illustrative purposes to show how to derive the underlying scaling hierarchy. More examples applied to geographic features for statistical mapping, map generalization, and cognitive mapping can be found in Jiang (2012a, 2012b) and Jiang et al. (2012). Note that these studies focused on deriving the scaling hierarchy rather than taking the hierarchical levels as a single index for characterizing the complexity of geographic features. Second, the pattern of far more small things than large things is shown in a very strict sense. For example, in this case the head is always less than 25 percent, while the tail is more than 75 percent. This may not be the case when dealing with geographic features (c.f., the case studies that follow). However, the unbalanced (or nonlinear) partition between the head and the tail hold remarkably the same for most geographic features. This is a striking signature of scaling of geographic space or features.

We have shown above that the iteratively derived means provide useful breaks for placing different segments into different hierarchical levels. Furthermore, the heads as minorities were distinguished from the tails as majorities. This is in a sharp contrast to conventional wisdom that the mean makes little sense to characterize data with a heavy-tailed distribution. Although conventionally the means cannot characterize the magnitude of data values, the means make perfect sense for deriving the inherent hierarchy.

**4. Case studies: Computing the ht-index of geographic features**
In this section, we will illustrate computation of the ht-index for a variety of geographic features at both country and city levels. At the country levels, we chose nightlight imagery and terrain heights for the first two case studies. At the city level, we utilized the degree of street connectivity for the third case study. Strictly speaking, nightlight is not a geographic feature, but it captures very well the pattern of human settlements.

**4.1 Ht-index of USA nightlight imagery**
As seen in the nightlight imagery (Figure 5), there appear to be far more dark pixels than light ones. The imagery contains 11,766,012 pixels. Each pixel has a light value ranging from 0 (darkest) to 63 (lightest). Our task is to apply the head/tail breaks to derive hierarchy, or classes in the imagery. Table 1 shows the results. Firstly, we can see that the average lightness of the 11,766,012 pixels is 7.5 (the first mean), which splits all the pixels into two unbalanced parts: 3,091,666 pixels (26 percent) above the first mean (in the head), and 8,674,346 pixels (74 percent) below the mean (in the tail, see the first row in Table 1). The average lightness of the 3,091,666 pixels in the head part is 23.0 (the second mean), which again splits the head into two unbalanced parts: 1,043,922 pixels in the head (34 percent) and 2,047,744 pixels in the tail (66 percent) (see the second row in Table 1). If we continue the similar partition for the 1,043,922 pixels, the second head is split into two well balanced parts: 50 percent above the mean and 50 percent below the mean (see the third row in Table 1). This third partition, however, is invalid, for it violates the condition of far more dark pixels than light ones. The two valid means lead to three classes, so the ht-index for the nightlight imagery is 3.

Table 1: Statistics for computing the ht-index of the night light imagery (Note: count = number of pixels; light*count = sum of individual light*count at each light level; # = number; % = percentage)

| Light | Count | Light*Count | Mean | # in head | % in head | # in tail | % in tail |
|---|---|---|---|---|---|---|---|
| 0-63 | 11,766,012 | 88,424,914 | 7.5 | 3,091,666 | 26% | 8,674,346 | 74% |
| 8-63 | 3,091,666 | 71,030,197 | 23.0 | 1,043,922 | 34% | 2,047,744 | 66% |
| 23-63 | 1,043,922 | 46,200,290 | 44.3 | 524,230 | 50% | 519,692 | 50% |

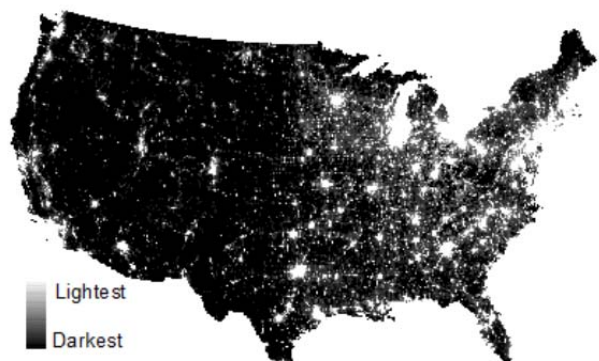

Figure 5: Nightlight imagery of USA in 2010

The violation of far more dark pixels than light ones implies that pixels with lightness greater than 23 are unlikely to exhibit a heavy-tailed distribution. Previous studies (Jiang 2012a, Jiang 2012b, Jiang et al. 2012) suggest that the percentages of the heads must be less than 40 percent. This condition can be relaxed for many geographic features, such as 50 percent or even more, if the head retains less than 40 percent in subsequent hierarchical levels. In other words, we further relax the condition for conducting head/tail breaks if most (rather than all) hierarchical levels meet the condition of far more small things than large ones. This will be shown in the third case study that follows. This relaxation of the condition is closely related to our dynamic view of fractals suggested in Section 2.2 and further discussed in Section 5.

**4.2 Ht-index of US terrain surface**

The United States terrain surface is far more complex than the imagery of nightlights. This is indeed true because natural phenomena are usually more complex than human-made phenomena. The digital elevation model (DEM) contains approximately 3 million pixels, with heights ranging from -147 to 4161 meters (Figure 6). The DEM is visualized by a stretched renderer using the particular color ramp. The overall perception of the visualized DEM is that there are far more low locations than high ones. This scaling hierarchy occurred at different places and scales (c.f., insets in Figure 6). For the sake of simplicity, we excluded negative height (or depth) and chose heights of at least 1 meter. Following the same procedures as in the first study, we derived an ht-index of 17. There were far more low-height pixels than high-height ones, and this pattern occurred 16 times (Table 2). As seen in Table 2, all the head percentages were less than 45 percent, indicating there were far more low-height pixels than high-height pixels at every hierarchical level. The computation of the ht-index was fairly simple and straightforward; one can simply rely on Excel to iteratively derive the means, and subsequently the ht-index equals the number of means plus 1.

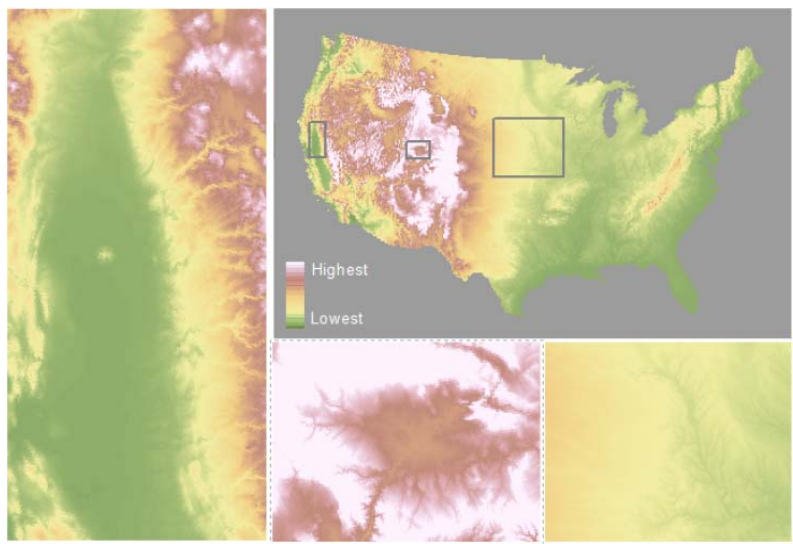

Figure 6: The DEM of the United States
(Note: The DEM is visualized by a stretched renderer using the particular color ramp with green as the lowest and white as the highest. The three insets indicate that the fractal property, or equivalently what looks irregular, wiggly or rough, occurred at different places and scales. The left inset indicates the full range of scales from the lowest to the highest, while the two bottom insets indicate respectively a highest scale range, and a lowest scale range.)

Compared to the nightlight imagery, the terrain surface has a very high ht-index of 17. This is understandable because (1) natural phenomena are usually more complex than human-made phenomena, and (2) the USA terrain surface is extremely heterogeneous and diverse. This case study implies that the ht-index might be a good indicator of complexity among different geographic features or different fractals.

Table 2: Statistics for computing the ht-index of the DEM(Note: count = number of pixels;
height*count = sum of individual height*count at each height; # = number; % = percentage).

| Height | Count | Height*Count | Mean | # in head | % in head | # in tail | % in tail |
|---|---|---|---|---|---|---|---|
| 1-4,161 | 2,929,517 | 2,336,290,278 | 798 | 1,124,266 | 38% | 1,805,251 | 62% |
| 798-4,161 | 1,124,266 | 1,781,726,590 | 1,585 | 497,151 | 44% | 627,115 | 56% |
| 1,585-4,161 | 497,151 | 1,037,436,307 | 2,087 | 200,212 | 40% | 296,939 | 60% |
| 2,087-4,161 | 200,212 | 497,334,830 | 2,484 | 76,586 | 38% | 123,626 | 62% |
| 2,484-4,161 | 76,586 | 218,834,200 | 2,857 | 30,959 | 40% | 45,627 | 60% |
| 2,857-4,161 | 30,959 | 98,139,789 | 3,170 | 12,933 | 42% | 18,026 | 58% |
| 3,170-4,161 | 12,933 | 44,133,166 | 3,412 | 5,620 | 43% | 7,313 | 57% |
| 3,412-4,161 | 5,620 | 20,158,052 | 3,587 | 2,400 | 43% | 3,220 | 57% |
| 3,587-4,161 | 2,400 | 8,915,043 | 3,715 | 1,022 | 43% | 1,378 | 57% |
| 3,715-4,161 | 1,022 | 3,892,555 | 3,809 | 411 | 40% | 611 | 60% |
| 3,809-4,161 | 411 | 1,596,335 | 3,884 | 166 | 40% | 245 | 60% |
| 3,884-4,161 | 166 | 655,118 | 3,947 | 62 | 37% | 62 | 37% |
| 3,947-4,161 | 62 | 247,943 | 3,999 | 26 | 42% | 36 | 58% |
| 3,999-4,161 | 26 | 104,940 | 4,036 | 10 | 38% | 16 | 62% |
| 4,036-4,161 | 10 | 40,737 | 4,074 | 4 | 40% | 6 | 60% |
| 4,074-4,161 | 4 | 16,429 | 4,107 | 1 | 25% | 3 | 75% |

**4.3 Ht-indices of urban streets**

For the third case study, we computed ht-indices of urban streets in terms of the degree of their connectivity. Cities are considered fractal, as Batty and Longley (1994) and Frankhauser (1994) showed, from the perspective of their shapes, land uses, and boundaries. However, we examined fractal cities from the perspective of street networks in terms of street hierarchies: far more less-connected streets than well-connected ones (Jiang et al. 2008, Jiang 2009, Sun 2013). The case study includes 48 cities, and their ht-indices vary from one to another (Table 3). Interestingly, all ht-indices computed ranged from 5 to 9. This indicates some universality of fractal property at the city level. The smallest city Vanersborg with only 141 streets has an ht-index of 5, while the largest city Stockholm with 15,175 streets has an ht-index 9. It should be noted, however, that there was no significant correlation between the city size and the ht-index. This implies that the ht-index is not simply concerned with the size of geographic features, but rather it deals with the underlying scaling or fractal property.

Table 3: Ht-indices of the 48 cities showing striking street hierarchies
(Note: h = ht-index, # = number, cities marked with * violate the 40 percent condition at least once,
the street hierarchies for the four highlighted cities are shown in Figure 7.)

| Cities | # streets | h | Cities | # streets | H | Cities | # streets | h |
|---|---|---|---|---|---|---|---|---|
| Boden | 186 | 5 | Kristianstad | 1051 | 7 | Skelleftea* | 947 | 7 |
| Boras* | 1902 | 7 | Landskrona* | 433 | 6 | Skovde* | 1054 | 7 |
| Borlange* | 2031 | 7 | Lidkoping* | 762 | 7 | Soderhamn* | 478 | 7 |
| Eskilstuna* | 761 | 6 | Linkoping | 5454 | 8 | Sodertalje* | 1305 | 7 |
| Falun | 2085 | 7 | Lulea* | 1306 | 7 | Stockholm | 15175 | 9 |
| Gavle | 3729 | 7 | Lund | 5861 | 9 | Sundsvall | 5261 | 7 |
| Goteborg* | 14918 | 8 | Malmo* | 5411 | 8 | Trelleborg | 625 | 6 |
| Halmstad* | 1460 | 7 | Motala* | 351 | 6 | Trollhattan* | 541 | 6 |
| Harnosand | 466 | 6 | Norrkoping* | 2585 | 8 | Uddevalla* | 910 | 7 |
| Helsingborg | 2291 | 7 | Norrtalje* | 446 | 6 | Umea* | 2833 | 8 |
| Jonkoping | 1772 | 7 | Nykoping* | 439 | 6 | Uppsala | 5556 | 9 |
| Kalmar* | 1631 | 7 | Orebro* | 1439 | 7 | Vanersborg | 141 | 5 |
| Karlskoga* | 511 | 6 | Ornskoldsvik* | 1261 | 7 | Varberg | 604 | 6 |
| Karlskrona | 1463 | 6 | Ostersund* | 3438 | 7 | Vasteras* | 1487 | 6 |
| Karlstad* | 3166 | 8 | Sandviken* | 846 | 7 | Vastervik | 316 | 6 |
| Kiruna* | 739 | 7 | Skara* | 297 | 6 | Vaxjo* | 3068 | 8 |

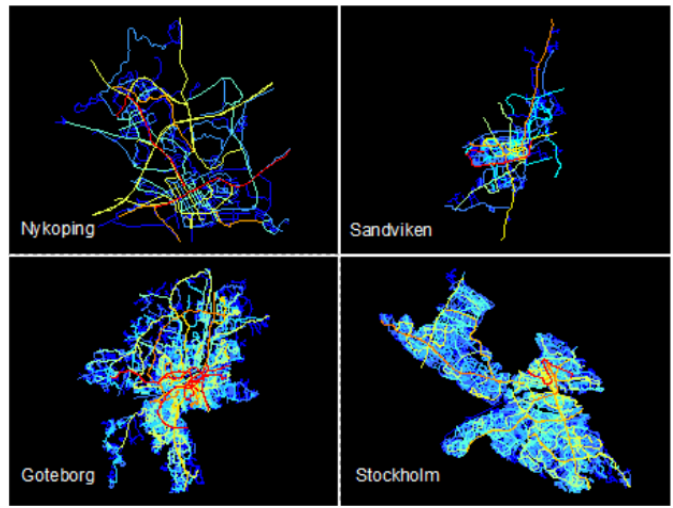

Figure 7: Street hierarchies visualized by spectral colors, with blue as the least connected and red as the most connected
(Note: The ht-indices of the four cities Nykoping, Sandviken, Goteborg, and Stockholm are respectively 6, 7, 8 and 9, while the city sizes are much more diverse ranging from hundreds to thousands.)

We have shown how the ht-index can characterize the complexity of geographic features, from nightlight imagery, to terrain surfaces, to city morphological structure. All these geographic features demonstrate the fractal or scaling property: in other words, there are far more small things than large ones. Thus the fractal is universal for most (if not all) geographic features. One peculiarity is whether or not the pattern of far more small things than large ones holds true at all or most hierarchical levels. This peculiarity may deserve further studies for it is related to different degree of fractal or complexity (c.f., further discussions in the next section). From the case studies, we have also seen that the means make a perfect sense in differentiating small and large things, and in deriving the ht-index.

**5. Implications of the ht-index**
We have seen in the case studies how the ht-index captures the underlying complexity of geographic features. The ht-index captures the inherent hierarchy of geographic features, which is of use for statistical mapping, map generalization, and cognitive mapping (Jiang 2012a, Jiang 2012b, Jiang et al. 2012). It fundamentally differs from the concept of fractal dimension, and captures what fractal dimension cannot. This section further discusses the ht-index, its relation to fractal dimension, and its implications for geography and beyond.

To a great extent, the ht-index complements fractal dimension in characterizing fractal complexity, and particularly that of geographic features. To further illustrate the relationship between the ht-index and fractal dimension, we adjusted the Koch curve a bit by increasing the segment length of the two middle segments. The adjustment increased the fractal dimension a small amount, due to the increment of the scaling ratio. This is seen in Figure 8, where from left to right, the degree of filled space has increased. On the other hand, the ht-index remains unchanged from left to the right, but only increases from top to bottom where fine scales are added. The underlying concept of fractals is scale or, more precisely, a wide range of scales involved in a fractal pattern or set. The wide range of scales implies heterogeneity, which is captured by both the ht-index and fractal dimension but from different perspectives. The ht-index expresses hierarchical levels for heterogeneous scales, while the fractal dimension expresses the degree of heterogeneity. In this regard, the ht-index provides a new measure for characterizing spatial heterogeneity of geographic features. This warrants some further studies to examine its effectiveness in measuring spatial heterogeneity.

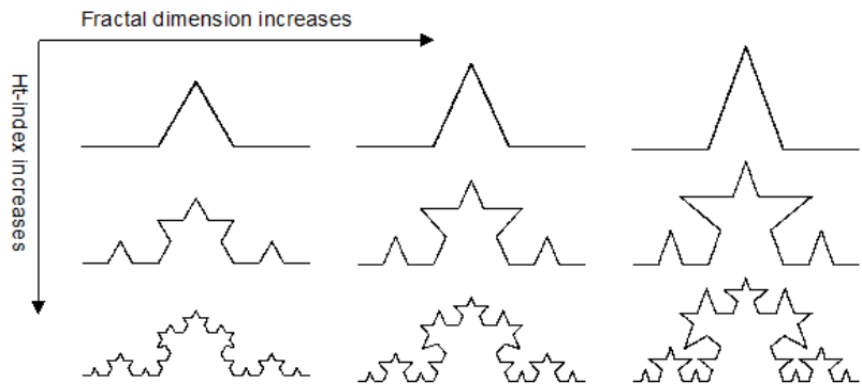

Figure 8: Both the ht-index and fractal dimensions, characterizing fractals from different perspectives
(Note: from the left to the right in each row, the length of two middle segments increases gradually, so does the scaling ratio. The increment of the length and the scaling ratio will lead to the increment of the fractal dimension from the left to the right. However, the fractal dimension remains unchanged from the top to the bottom in each column. Obviously, there is a change from the top to the bottom, i.e., the gradually added fine scales, and this change is captured by the ht-index.)

The ht-index provides a dynamic view of examining fractals, or geographic forms and processes in particular. Instead of the fractal as a clear-cut concept, we believe it is a fuzzy concept, ranging from less to more fractal. As the ht-index increases, more fine structures are added (e.g., Koch curve) or removed (e.g., Cantor set), becoming more fractal. This is exactly what the ht-index captures, yet fractal dimension misses. This dynamic view of fractals can be seen from another perspective. A fractal is characterized by a power-law distribution relating change in scale ($r$) with change in details ($N$), while a less fractal is described by a power law-like distribution such as the lognormal or exponential. More importantly, geographic features are continuously evolving and developing. Something that is less fractal now could be fractal tomorrow. Something that is fractal now could be more fractal tomorrow. This is the dynamic view we advocate to better understand geographic forms and processes, and their relationships.

Mandelbrot's (1967) major contribution lies in his relaxed definition of self-similarity, from the initial strict sense (such as Koch curves) to the statistical sense (such as coastlines). We further relaxed Mandelbrot's definition by moving away from a power law distribution to a power law-like distribution such as the lognormal or exponential, as long as the pattern of far more small things than large ones holds true. In fact, this condition has been further relaxed from having to hold for all hierarchal levels (Jiang 2012) to holding for most levels, as seen in the third case study. We believe that the fractal is an idealized status, and many geographic features develop toward it, as our dynamic view suggests.

We rank and differentiate between large and small things, and iteratively continue the ranking and differentiating processes to derive the ht-index. Our purpose is to better understand the underlying scaling structure, and how things of different scales form a hierarchy. In this regard, the concept of the ht-index can be applied to any fields or domains for characterizing the complexity of fractal patterns or sets. All scales or all things of different sizes are collectively essential for the scaling hierarchy as a whole. However, from an individual viewpoint, large things could be more important than small things. Taking an urban street network as an example, all streets including short or less-connected streets are essential for the street hierarchy (Jiang 2009), and our daily life substantially relies on short or less-connected streets. However, for the purpose of map generalization and due to the limitation of space (Jiang et al. 2012), we tend to consider the large things to be more important, and eliminate small things while retaining large ones.

**6. Conclusion**
This paper developed the ht-index to quantify the fractal or scaling structure of geographic features based on the notion of far more small things than large ones. Moving away from the strict definition of fractal dimension, we provided a fairly simple solution for characterizing the fractal or scaling structure. This was mainly motivated by the dilemma we face: fractal or scaling is said to be the norm for geographic features, but they hardly qualify as fractals according to the traditional definition of fractals. We defined fractal or scaling as a recurring structure of far more small things than large ones. The ht-index captures the hierarchical levels of the recurring structure.

The ht-index is not intended to replace fractal dimension, but it is an alternative for capturing the detailed aspect of fractals. Geographic features with a higher ht-index are more complex, heterogeneous, mature, and/or natural. In this regard, the ht-index is important for measuring heterogeneity, particularly spatial heterogeneity. A higher ht-index also implies more information (Jiang 2012b), for geographic features with a higher ht-index tend to be more heterogeneous and more diverse. Although the ht-index is inspired by fractal structure of geographic features, and developed for characterizing the universal structure, its applications are not just limited to geography. We believe the ht-index can be applied to a broad range of fractals in nature and society. This warrants further studies, particularly in other disciplines where fractals are widely observed.

**Acknowledgement**
We are grateful to the anonymous reviewer and Mei-Po Kwan for their constructive comments that significantly helped improve the quality of this paper. We also would like to thank Benny Jiang for several conversations and discussions during which we corrected one critical mistake in an earlier version of this paper. However, any possible shortcomings remain our responsibility.